\documentclass[fleqn,usenatbib]{mnras}

\usepackage[T2A]{fontenc}
\usepackage[utf8]{inputenc}
\usepackage[english]{babel}
\usepackage{hyperref}

\usepackage{graphicx}	% Including figure files
\usepackage{subfigure}
\usepackage{amsmath}	% Advanced maths commands
\usepackage{amssymb}	% Extra maths symbols
\hypersetup{draft}
\usepackage{multirow}

\usepackage{xcolor}
\definecolor{mygray}{gray}{0.6}
\definecolor{myblue}{RGB}{20, 20, 200}
\definecolor{myred}{RGB}{200, 20, 20}

\title[New LBVs in NGC\,1156]{Search for LBVs in the Local Volume galaxies: study of two stars in NGC 1156}

\author[Y. Solovyeva et al.]{
Y. Solovyeva,$^{1}$\thanks{E-mail:solovyeva@sao.ru}
A. Vinokurov,$^{1}$
N. Tikhonov,$^{1}$
A. Kostenkov,$^{1}$
K. Atapin,$^{2}$
A. Sarkisyan,$^{1}$
\newauthor
A. Moiseev,$^{1}$
S. Fabrika,$^{1}$
D. Oparin,$^{1}$
A. Valeev$^{1}$
\\
$^{1}$Special Astrophysical Observatory, Nizhnij Arkhyz 369167, Russia\\
$^{2}$Sternberg Astronomical Institute, Lomonosov Moscow State University, Universitetskij Pr. 13, Moscow 119992, Russia\\}

\date{Accepted XXX. Received YYY; in original form ZZZ}

\pubyear{2022}

\begin{document}
\label{firstpage}
\pagerange{\pageref{firstpage}--\pageref{lastpage}}
\maketitle

\begin{abstract}

We continue the search for luminous blue variables (LBVs) in Local Volume galaxies in order to study their fundamental parameters. In this paper, we report the discovery of two new LBVs in the dwarf irregular galaxy NGC\,1156. Both stars exhibit spectral variability simultaneously with strong brightness variations: $\Delta \text{R}_c = 0.84 \pm 0.23^m$ for J025941.21+251412.2 and $\Delta \text{R}_c = 2.59 \pm 0.10^m$ for J025941.54+251421.8. The bolometric luminosities of the stars are in the range of L$_\text{Bol} \approx (0.8  - 1.6) \times 10^6$~L$_\odot$. These values are corrected for reddening A$_\text{V} \approx 0.9$ and are given for the distance to the galaxy D=7.0$\pm$0.4~Mpc, which we have measured by the TRGB method. Both stars are above the Humphreys-Davidson limit in the region of relatively low temperatures, T$_\text{eff} \lesssim 10$~kK on the temperature-luminosity diagram. J025941.54+251421.8 had a temperature below the hydrogen ionisation threshold at maximum brightness, exhibiting behaviour very similar to that of the known LBV R71 during its 2012 outburst. We have estimated the masses of the detected LBVs and studied the properties of their stellar environment. We discuss our results within the framework of both a single star and a binary system evolution scenario for LBVs.

\end{abstract}


%%%%%%%%%%%%%%%%%%%%%%%%%%%%%%%%%%%%%%%%%%%%%%%%%%

%%%%%%%%%%%%%%%%% BODY OF PAPER %%%%%%%%%%%%%%%%%%

\section{Introduction}

Luminous blue variables (LBVs) are massive stars ($M\geq25M\odot$, \citealt{Humphreys16}) located at the top of the Hertzsprung-Russell diagram and characterised by irregular variability on different time scales \citep{vanGenderen01}. The spectra of LBVs are somewhat similar to those of other luminous stars, such as B[e]-supergiants, warm hypergiants, Of/late-WN stars and others \citep{Humphreys14}. The similarity of observational manifestations complicates the identification of LBVs among similar stars; however, their main distinguishing feature is the S Dor type variability, which consists of the stellar magnitude varying by up to $\approx2.5^m$ simultaneously with its spectrum. During the cycle of S Dor type variability, the bolometric luminosity of LBVs remains approximately constant. Nevertheless, it has been shown in a number of works (for example, \citealt{Groh09, Clark09, Lamers95}) that some LBVs exhibiting S Dor type variability demonstrated a decrease in their bolometric luminosity during the visual brightness maximum, which is probably caused by the energy losses due to the expansion of the outer layers of the star. In addition, some LBVs also show $\eta$ Car type variability \citep{Humphreys99}, which consists of extreme brightness changes in the form of giant eruptions (with an amplitude $\gtrsim 2.5^m$). The reasons for the variability of both types are not completely clear. These stars are often surrounded by compact dust envelopes formed during the ejection of stellar matter. Using LBVs and LBV candidates (cLBVs) in the Magellanic Clouds as an example, \citet{Agliozzo21} have shown that such objects are an important source of dust in galaxies.

The traditional view considers LBVs as a transitional stage between single O-stars with an initial mass $\gtrsim$25 M$\odot$ and Wolf-Rayet (WR) stars (the Conti scenario, \citealt{Conti75, Conti84}) implying that the transition is from the core hydrogen burning to the core helium burning. The stellar wind and outbursts at the LBV stage play a decisive role in the removal of the hydrogen envelope necessary for the formation of a WR-type star \citep{Groh14}. Some authors have suggested that LBVs may be the immediate precursors of core collapse supernovae \citep{Groh13}, and this idea is supported by several studies \citep{Trundle08, Andrews21}. 

An alternative explanation of the LBV phenomenon considers these stars as the result of close binary evolution. One confirmation of this idea is the discovery of the spatial isolation of LBVs with respect to massive O stars \citep{Smith15, Smith16}, and the degree of this isolation is much greater than the single massive star evolution models allow. However, the reliability of these conclusions strongly depends on the sampling method used for the stars under consideration, as pointed out by \citet{Humphreys16}. Nevertheless, modelling young star clusters and their dissolution \citep{Aghakhanloo17} also led the authors to the conclusion that the standard model of single star evolution is largely inconsistent with the stellar environment of LBVs, and that LBVs are ``rejuvenated stars'' in binary systems that accreted mass from a more massive neighbour. The probable binarity of up to 70\%\ of galactic LBVs and cLBVs is demonstrated in a recent paper by \citet{Mahy22} based on the results of long-term spectroscopy and interferometry of these stars. However, the question of whether LBVs are an evolutionary stage of a single massive star or a product of binary star evolution is still under discussion.

It is worth noting that all studies aimed at explaining the LBV phenomenon are based on relatively small samples, since only a little more than 40 LBVs and about 100 cLBVs are currently known \citep{Richardson18}. Therefore, to obtain more statistically significant results, a search for new LBV stars is necessary. Our Galaxy and the nearest galaxies of the Local Group including M\,33, M\,31 have already been studied fairly well for the presence of LBV stars, and a search for these stars performed by our group in galaxies outside the Local Group looks more promising. In this paper, we present the search results for the NGC\,1156 galaxy.

The paper is organised as follows: Section 2 reports the details of the observations, data reduction, and measuring the distance to the galaxy; Section 3 presents the results of the spectroscopy, photometric analysis, and estimating stellar parameters and ages of the surrounding stars. In section 4, we present the classification of the objects and discuss their initial masses in the context of single and binary star evolution.

%Fig.1
\begin{figure*} 
\includegraphics[angle=0, width=0.9\linewidth]{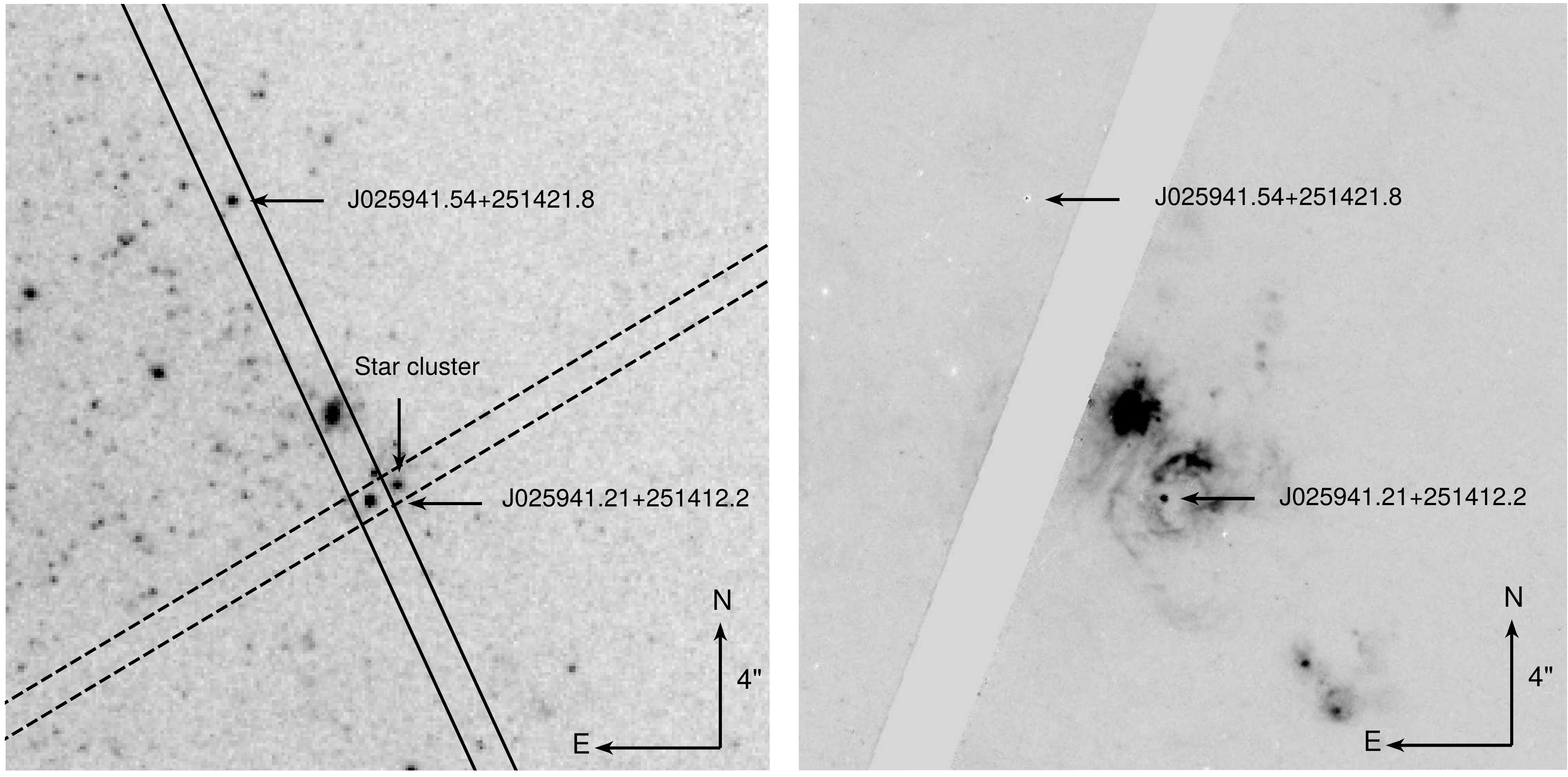}
\caption{\textit{left panel:} Image obtained with the HST/WFPC2 in the F814W filter. The 1.2\arcsec and 1.0\arcsec slits are shown by solid and dashed lines, respectively. \textit{right panel:} The HST/ACS/WFC/F658N image with subtracted continuum contribution (ACS/WFC/F625W).} 
\label{Fig1} 
\end{figure*}

\section{OBSERVATIONS AND DATA REDUCTION}

The NGC\,1156  is a dwarf irregular Magellanic-type galaxy which has a relatively high star formation rate\footnote{According to the catalog of Local Volume galaxies https://relay.sao.ru/lv/lvgdb/}: 0.3$M_
\odot\,\mbox{yr}^{-1}$. The distance to the galaxy is known with poor accuracy, and its measurement was carried out only by a few methods: the estimates vary from 3.78 Mpc (\citealt{Bottinelli1984}, the Tully-Fisher relation) to 8.13 Mpc (\citealt{Kim2012}, the brightest stars of the galaxy). Since the accuracy of the distance measurements plays a decisive role in determining fundamental stellar parameters, we decided to estimate it ourselves by using the TRGB method (tip of the red-giant branch, \citealt{Lee1993}), which employs the luminosity of the brightest stars at the red-giant branch as a distance indicator and provides an accuracy of about 0.1$^m$ (\citealt{Lee1993}; see Sec.~\ref{subsec:dist}). 

We utilised archival wide- and narrow-band (H$\alpha$) images obtained with the Hubble Space Telescope (HST). The point sources in the H$\alpha$ images associated with bright blue stars were flagged as candidates for detailed studies. This method has proven its effectiveness, our previous results are published in a number of papers \citep{Solovyeva19,Solovyeva20, Solovyeva21}. Checking HST images of NGC\,1156, we have found one LBV candidate: J025941.21+251412.2 (Fig.~\ref{Fig1}). The second LBV candidate, J025941.54+251421.8, was discovered during spectral observations by bright H$_\alpha$ emission in its spectrum. Both objects are located in regions of active star formation. Their status was clarified by spectroscopy and photometry.

\subsection{Spectroscopy}

The stellar spectra were obtained with the 6-m Big Telescope Alt-azimuth (BTA) of SAO RAS  using the SCORPIO\citep{Afanasiev05} and SCORPIO-2 \citep{AfanasievSco2} multi-mode prime focus reducers operating in the long slit mode. The observation log is given in Table~\ref{tab:spec}. In 2013--2020 yrs we oriented slit in the position angle $PA=24\degr$ to get simultaneously  spectra of the both targets Fig.~\ref{Fig1}). In 2021 yr the slit was oriented in $PA=124\degr$ in order to separate the light contribution of the LBV candidate and star cluster (see below). 
The spectral data were processed in a standard way using the \textsc{long} context in the \textsc{midas} environment. The extraction of one-dimensional spectra was performed using the \textsc{spextra} program designed to deal with spectra in crowded stellar fields \citep{Sarkisyan17}. 

\begin{table*}
\caption{BTA spectral observation log.}
\begin{tabular}{ccccccc}
\hline\hline 
Date& Spectrograph/grism & Total exp., & Slit width & Spectral  & Spectral  & Seeing\\  
& & s. &  resolution,\AA & range, \AA  &  & \\  
\hline
2013/10/31 & SCORPIO/VPHG1200R & 1850 & 1.2\arcsec & 5.0 & 5700-7400 & 1.2\arcsec \\
2013/12/31 & SCORPIO/VPHG1200G & 4800 & 1.2\arcsec & 5.0 & 3900-5700 & 1.0\arcsec \\
2017/09/19 & SCORPIO/VPHG1200G & 1200 & 1.2\arcsec & 5.0 & 3900-5700 & 1.6\arcsec \\
2019/10/23 & SCORPIO/VPHG1200B & 3600 & 1.2\arcsec & 5.5 & 3600-5400 & 1.3\arcsec \\
2020/09/23 & SCORPIO/VPHG1200B & 1800 & 1.2\arcsec  & 5.5 & 3600-5400 & 1.5\arcsec \\
2020/09/23 & SCORPIO/VPHG1200R & 1800 & 1.2\arcsec & 5.0 & 5700-7400 & 1.1\arcsec \\ 
2021/10/11 & SCORPIO-2/VPHG1200@540 & 2700 & 1.0\arcsec & 5.5 & 3650-7250 & 1.1\arcsec \\ 
\hline
\end{tabular}
\label{tab:spec}
\end{table*}

\subsection{Imaging}
\label{subsec:imaging}

We have carried out multiple observations of NGC\,1156 with the 2.5-m telescope of the Caucasian Mountain Observatory of SAI MSU \cite[2.5-m CMO,][]{KGO2020} and with the BTA since the discovery of both stars in 2013. The 2.5-m telescope observations were performed with NBI $4K\times4K$ optical CCD camera, whereas at the BTA we used the same SCORPIO focal reducer in the direct image mode.
We have also analysed archival data of Wide Field Channel and High-Resolution Channel of the Advanced Camera for Surveys (ACS/WFC and ACS/HRC respectively), and the Wide Field and Planetary Camera 2 (WFPC2) of HST\footnote{The data were taken from MAST archive \url{https://archive.stsci.edu/}}. Observation dates and filters are listed in Tables~\ref{tab:hst} and \ref{tab:phot}.

We performed point-spread function (PSF) and aperture photometry on the HST images. The PSF  photometry (only for WFPC2 data) was done on the preprocessed c0f images using \textsc{hstphot}\,1.1 \citep{Dolphin00}. For the ACS data, the aperture photometry was performed on drс images using the \textsc{apphot} package in \textsc{iraf}. Stellar fluxes were measured in a 3-pix aperture, which corresponds to 0.15\arcsec\ and 0.075\arcsec\ for WFC and HRC, respectively. The sky background was determined in annular apertures with an inner radius of 0.25\arcsec\ and an outer radius of 0.4\arcsec. Aperture corrections were obtained based on the photometry of 14 (ACS/WFC) and 18 (ACS/HRC) single bright stars. The final stellar magnitudes in the VEGAMAG system are presented in Table~\ref{tab:hst}.

For comparison of the photometry results with ground-based observations, the measured HST magnitudes were converted to the Johnson-Cousins system using the PySynphot package (Table~\ref{tab:phot}). As a model spectrum for this conversion we used a power law with spectral indices derived from the fluxes in two adjacent filters.

\begin{figure} 
\center{\includegraphics[angle=0, width=0.8\linewidth]{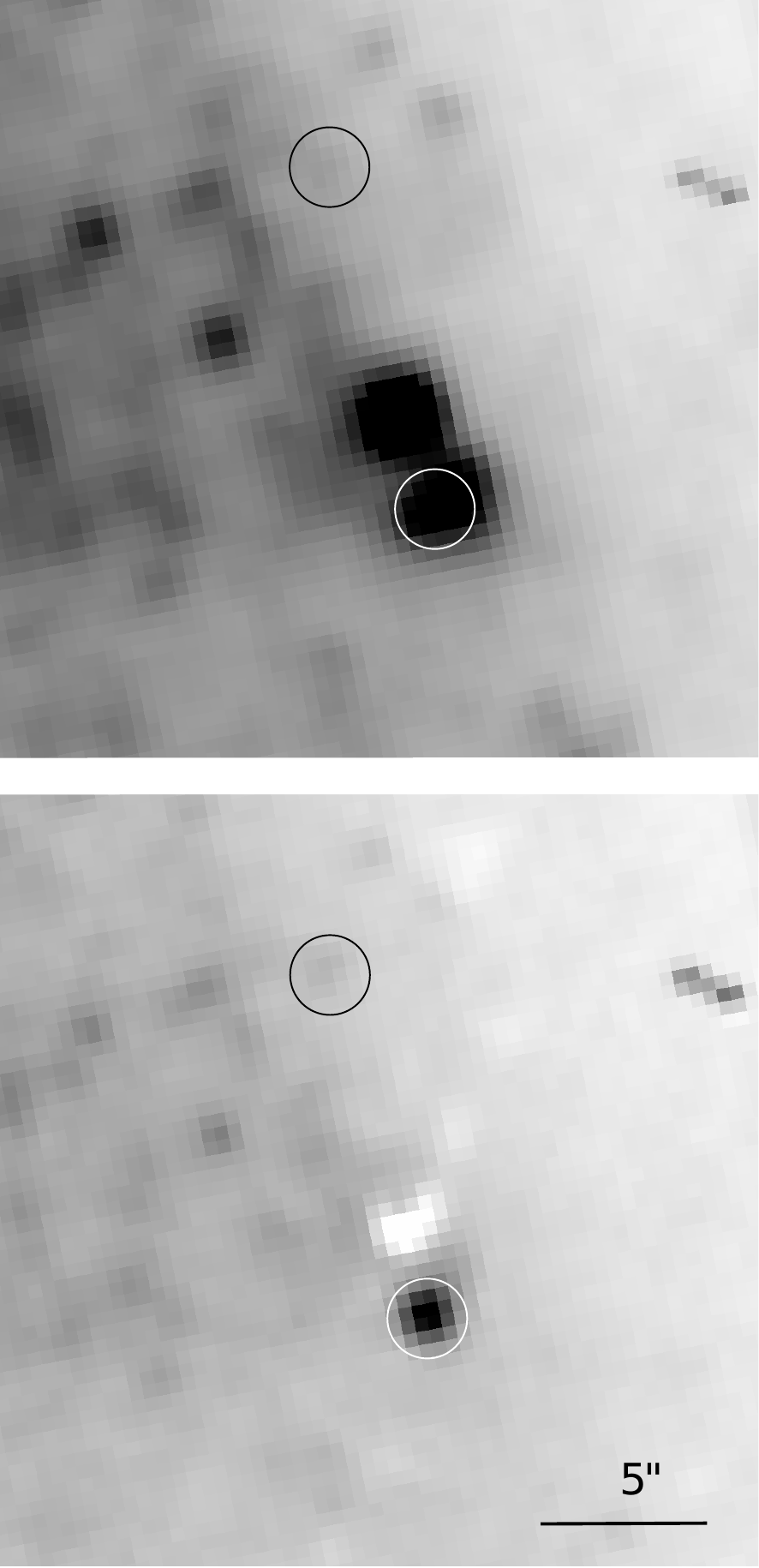}}
\caption{\textit{Top panel}: 
R-band image of the region including J025941.21+251412.2 (white circle) and J025941.54+251421.8 (black circle) obtained on 2021/10/11 with the BTA. \textit{Bottom panel}: The same image after the subtraction procedure (see text for details).
} 
\label{Fig2} 
\end{figure}

Pre-processing of raw images obtained with the ground-based telescopes (bias subtraction, flat-fielding and removal of cosmic ray events) was performed in the \textsc{midas} environment. The seeing variations from night to night were within FWHM=0.9\arcsec--1.3\arcsec, which is still insufficient for resolving the targets from neighbouring sources even when using the PSF photometry method. Moreover, whereas in the case of J025941.54+251421.8 its nearest environment contributes less than 30\%\ to the total flux in an aperture equal to the stars' FWHM (regardless of the filter), the neighbours of J025941.21+251412.2 contribute a flux comparable to that of the object itself.
We have (almost completely) subtracted this contribution using the smoothed HST images as maps of the sky background in the areas near the studied objects. We did the following. Both the studied stars and the 13 comparison stars that were used for absolute calibrations were manually removed (masked) from the HST images in the ACS/WFC/F606W and F814W filters (taken on 2019/12/24) whose pass bands are close to those of the Johnson-Cousins V, R$_c$ and I$_c$. The resulting images were then smoothed using a Gaussian filter with the width corresponding to the seeing in a particular ground-based observation and subtracted pixel by pixel: the F606W image from the V and R$_c$ data, the F814W one from the I$_c$ data. The level between each pair of images (minuend and subtrahend) was adjusted with a scaling factor, which was chosen in such way as to achieve the most complete subtraction of all the objects except the masked ones.

Despite the incomplete coincidence of the pass bands of the Johnson-Cousins and HST filters, our method allowed us to significantly reduce the flux contribution of the neighbouring stars and to minimise the flux measurement ambiguities. The fluxes in the resulting images were measured by aperture photometry using the \textsc{apphot} package. The remaining background (associated with the night sky brightness and other scattered light in the ground-based observations) was estimated in annular apertures around the objects with the inner and outer radii equal to 2.5\arcsec\ and 4\arcsec\ (slightly varied depending on the telescope). Fig.~\ref{Fig2} shows an example of the same frame before and after the subtraction; the photometry results are presented in Table~\ref{tab:phot}.

\begin{table}
\caption{Results of the HST photometry. All magnitudes are given in the VEGAMAG system.}
\begin{tabular}{lcc} \hline\hline 
\centering
Camera & Date & Filter and magnitude  \\ \hline
\multicolumn{3}{c}{J025941.21+251412.2} \\ \hline
WFPC2 & 2001/07/07 & F300W=$20.31\pm0.03$\\
 &  & F814W=$19.09\pm0.02$\\\hline
ACS/WFC & 2004/03/18 & F625W=$19.49\pm0.02$\\
 &  & F658N=$18.30\pm0.03$\\
\hline
\multicolumn{3}{c}{J025941.54+251421.8} \\ \hline
WFPC2 & 2001/07/07 & F300W=$21.80\pm0.07$\\
 &  & F814W=$19.60\pm0.03$\\\hline
ACS/WFC & 2004/03/18 & F625W=$19.64\pm0.02$\\
 &  & F658N=$19.30\pm0.03$\\\hline
ACS/HRC & 2005/09/05 & F330W=$21.54\pm0.04$\\
 &  & F435W=$20.77\pm0.05$\\
 &  & F550M=$20.04\pm0.05$\\
 &  & F658N=$19.66\pm0.05$\\
 &  & F814W=$19.36\pm0.04$\\
 \hline
ACS/WFC & 2019/12/24 & F606W=$22.25\pm0.03$\\
 &  & F814W=$22.06\pm0.03$\\
\hline
\end{tabular}
\label{tab:hst}
\end{table}

\begin{table*}
\caption{Results of ground-based photometry and  comparission with HST data. The columns show the instruments, dates and observed stellar magnitudes (not corrected for reddening). Magnitudes measured from the HST data were converted to the Johnson-Cousins filters. All magnitudes are given in the VEGAMAG system.}
\begin{tabular}{lcccc} \hline\hline 
\centering
Telescope & Date & V, mag & R$_\text{c}$, mag & I$_\text{c}$, mag \\ \hline
\multicolumn{5}{c}{J025941.21+251412.2} \\ \hline
WFPC2 & 2001/07/07 & --- & --- & $19.59\pm0.03$ \\
ACS/WFC & 2004/03/18 & --- & $19.47\pm0.03$ & --- \\
BTA & 2013/10/31 & --- & $20.20\pm0.23$& --- \\
2.5m CMO & 2018/09/20 & $19.91\pm0.04$ & $19.36\pm0.02$ & ---\\
BTA & 2019/10/23  & --- & $19.40\pm0.14$ & --- \\
2.5m CMO & 2020/09/10 & --- & --- & $19.34\pm0.02$\\
BTA & 2020/11/18 & --- & $19.34\pm0.13$ & --- \\
BTA & 2021/10/11 & --- & $19.41\pm0.03$ & ---  \\
\hline
\multicolumn{5}{c}{J025941.54+251421.8} \\ \hline
WFPC2 & 2001/07/07 & --- & --- & $19.03\pm0.03$ \\
ACS/WFC & 2004/03/18 & --- & $19.63\pm0.03$ & --- \\
ACS/HRC & 2005/09/05 & $20.07\pm0.05$ & $19.77\pm0.06$ & $19.43\pm0.04$ \\
BTA & 2013/10/31 & --- & $21.09\pm0.06$& --- \\
2.5m CMO & 2018/09/20 & $22.77\pm0.13$  & $22.22\pm0.10$ & ---\\
ACS/WFC & 2019/12/24 & $22.30\pm0.05$ & $22.19\pm0.03$ & $22.05\pm0.03$\\
\hline
\end{tabular}
\label{tab:phot}
\end{table*}

\subsection{Distance to the NGC\,1156 galaxy}
\label{subsec:dist}

We used the archival HST images obtained with the ACS camera in 2019 to perform the photometry of the NGC\,1156 stars and to measure the distance. Stellar photometry was carried out in a standard way similar to that described earlier in \citet{Tikhonov2019} using the \textsc{daophot}~II package \citep{Stetson1987, Stetson1994}. Stars were selected by the ``CHI'' and ``SHARP'' parameters that determine the photometric profile shape of each measured star \citep{Stetson1987}, allowing one to filter out non-stellar objects (star clusters, distant or compact galaxies). The profiles of non-stellar objects differ from those of isolated stars chosen as standard, so this allowed us to carry out such a selection for all the stars on the list.

The galaxy NGC\,1156 is located at the centre of the HST images and occupies the entire frame. Since the presence of bright stars distorts the position of the branch of red giants, which are fainter stars, we considered only the stars at the galaxy edge for these measurements, where the number of bright stars is insignificant. Red giants are clearly visible in the HST images of the galaxy periphery, so the TRGB method \citep{Lee1993} can give reliable distance estimates. We used the Sobel function \citep{Madore1995} to find the position of the TRGB jump which defines the boundary of the red giant branch. The maximum of this function indicates a sharp change in the number of stars observed at the red giant branch boundary.

\begin{figure} 
\center{\includegraphics[angle=0, width=0.75\linewidth]{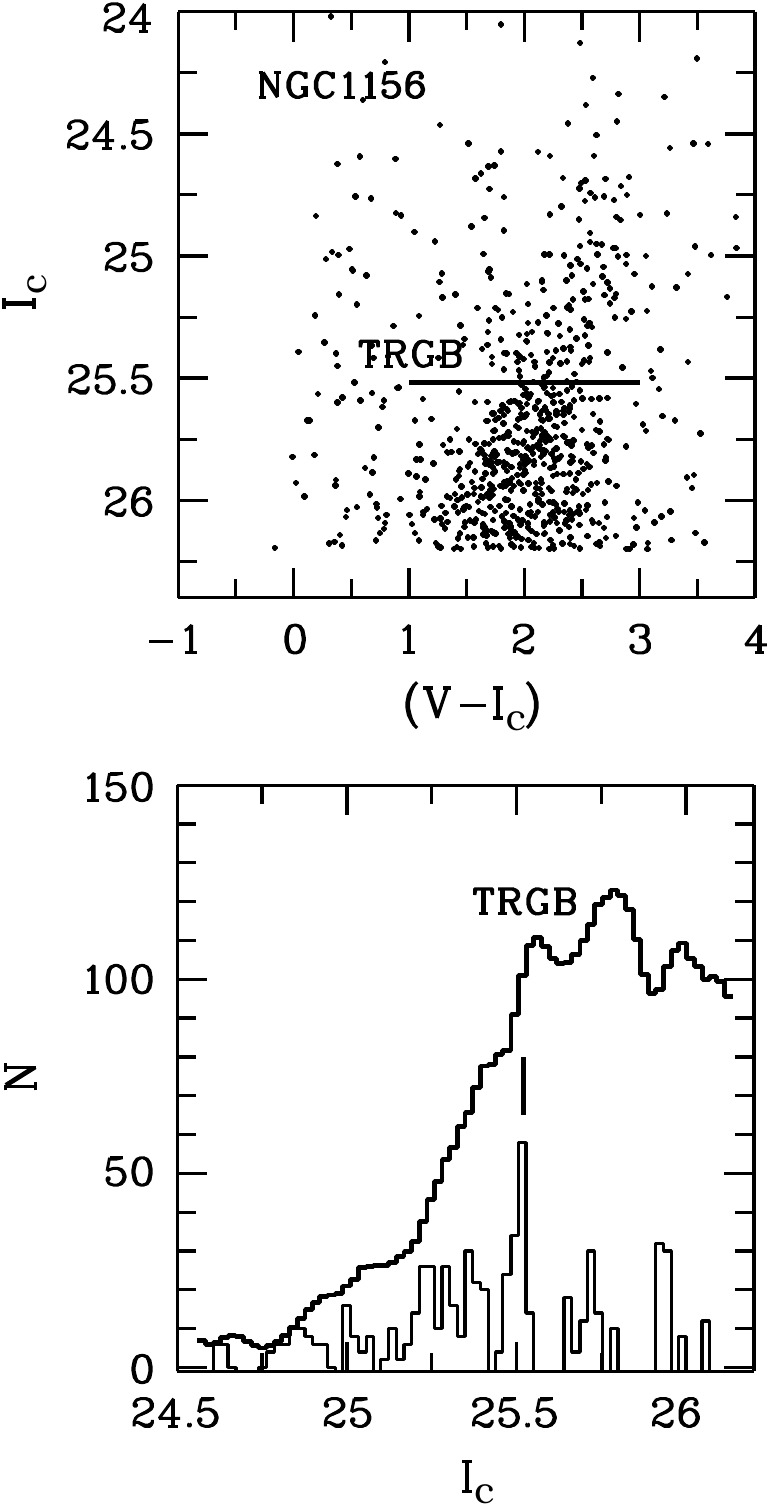}}
\caption{
The colour--magnitude diagram for the NGC\,1156 periphery stars ({\it left}) and the luminosity function of red giants and AGB stars ({\it right}). The maximum of the Sobel function indicates the position of the TRGB jump at I$_\text{c}$ = 25.52$^m$.
} 
\label{Fig3} 
\end{figure}

Fig.~\ref{Fig3} shows the Colour-Magnitude diagram (CMD)\footnote{The measured magnitudes were converted from the F606W and F814W filters to comparable Johnson-Cousins V and I$_\text{c}$, respectively.} of the NGC\,1156 periphery stars and the luminosity function of the red giants and asymptotic giant branch stars, where the Sobel function is marked by a thin line. The maximum of this function determines the position of the TRGB jump at I$_\text{c} = 25.52^m$. The colour indices of the red giant branch at the M$_\text{I}\text{c}=-3.5^m$ level and at its top are $(\text{V}-\text{I}_\text{c})_{-3.5} = 1.73^m$ and $(\text{V}-\text{I}\text{c})_{\text{TRGB}} = 1.90^m$, respectively. This allows us to determine the distance to NGC\,1156 and the metallicity of the red giants at the periphery of NGC\,1156 based on the relations from \citet{Lee1993}: (m-M) = 29.23$^m$, D = 7.01$\pm$0.40 Mpc, [Fe/H] = -1.31. The interstellar reddening towards the NGC\,1156 galaxy ($\text{A}_\text{V} = 0.614^m$ and $\text{A}_\text{I}\text{c} =0.337^m$) was taken from \citet{Schlafly2011}.

As was noted above, the accuracy of the method proposed in \citet{Lee1993} is about $0.1^m$. The total additional error $0.03^m$ of the distance measurement consists of the photometry error of TRGB stars, the PSF profile measurement error and the error of magnitude conversion between photometric systems. This defines the accuracy of our measurements: $(m-M) = 29.23\pm0.13$ or $D = 7.01\pm0.40$~Mpc.

\section{Results}
\label{res}

\subsection{Spectroscopy and photometry}
\subsubsection{J025941.21+251412.2}

\begin{figure*} 
\begin{center}
\includegraphics[angle=270, width=0.99\linewidth]{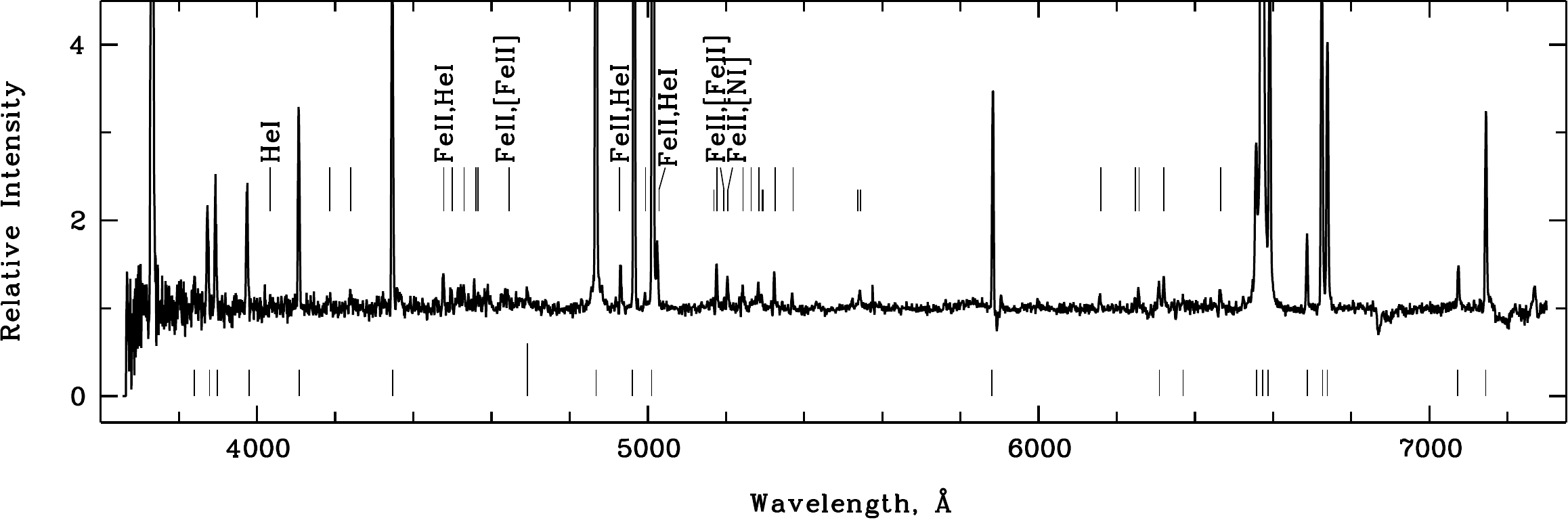}\\
\caption{Normalised spectrum of J025941.21+251412.2 obtained in 2021. The unlabelled long and short ticks above the spectrum represent the \ion{Fe}{ii} and [\ion{Fe}{ii}] lines, respectively. The short ticks under the spectra indicate the main lines of the nebula around the object, and the unlabelled long tick shows \ion{He}{ii} emitted by the nearest star cluster.} 
\label{Fig4} 
\end{center}
\end{figure*}

The object spectrum of the best spectral resolution in the widest wavelength range obtained in 2021 is shown in Fig.~\ref{Fig4} together with the identified lines. The spectrum contains hydrogen lines with broad wings and numerous \ion{Fe}{ii} and [\ion{Fe}{ii}] lines, which are characteristic of many types of massive stars with a strong gas outflow. Bright ionised iron lines indicate relatively low temperatures of the stellar wind (10-15 kK), however, the spectrum also shows \ion{He}{ii} $\lambda$4686 emission, which has to correspond to a higher degree of gas ionisation. We suppose that this line may belong to a young star cluster located nearby (see below); the narrow components of the hydrogen Balmer lines, helium \ion{He}{i}, forbidden lines [\ion{O}{iii}], [\ion{O}{i}], [\ion{N}{i}] , [\ion{N}{ii}], [\ion{S}{ii}] and [\ion{Ar}{iii}] are, apparently, partially or completely formed in the gas of the nearest \ion{H}{ii} region.

\begin{figure}
\begin{center}
\includegraphics[angle=270, width=0.9\linewidth]{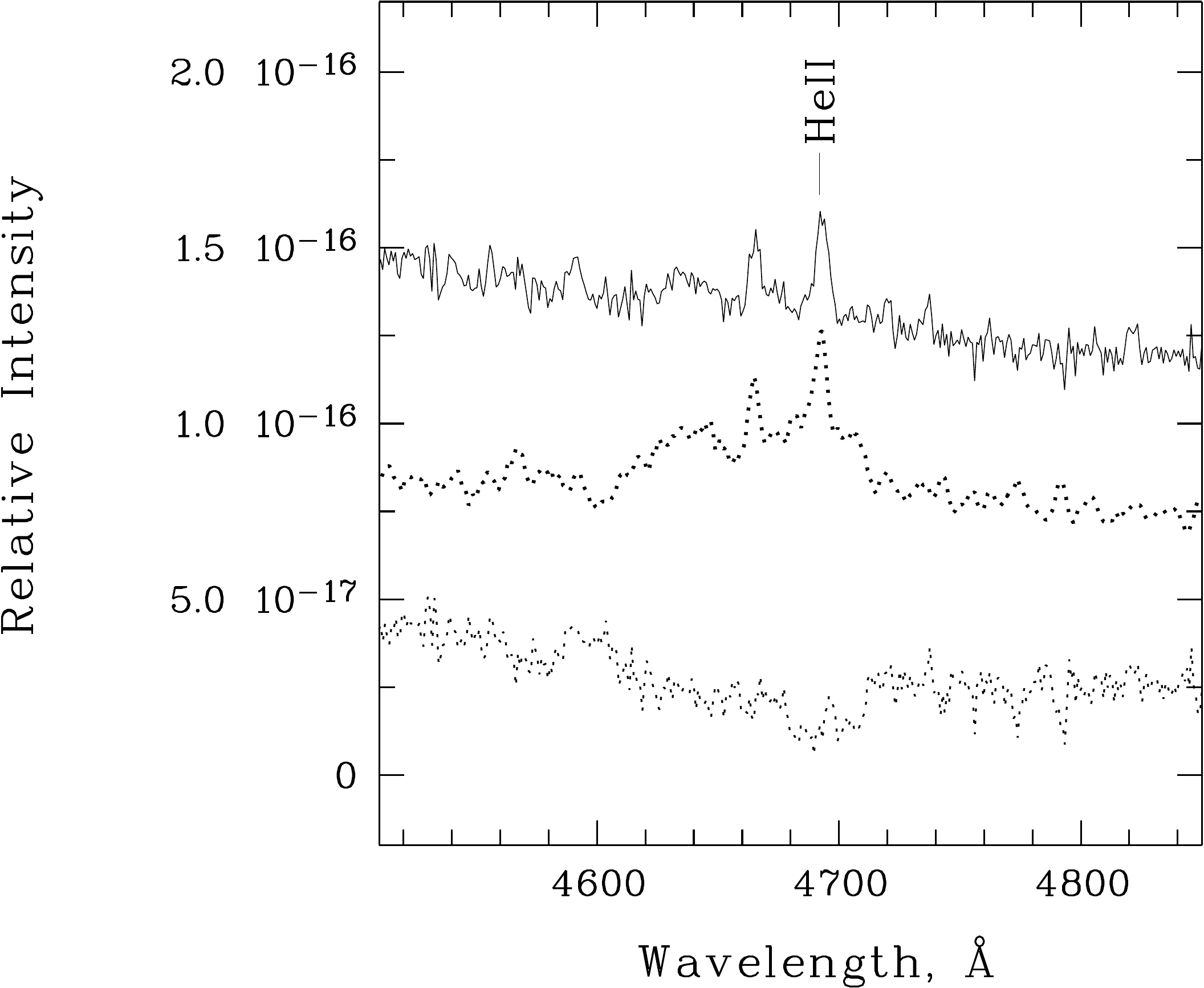}\\
\vspace{5pt}
\includegraphics[angle=270, width=0.9\linewidth]{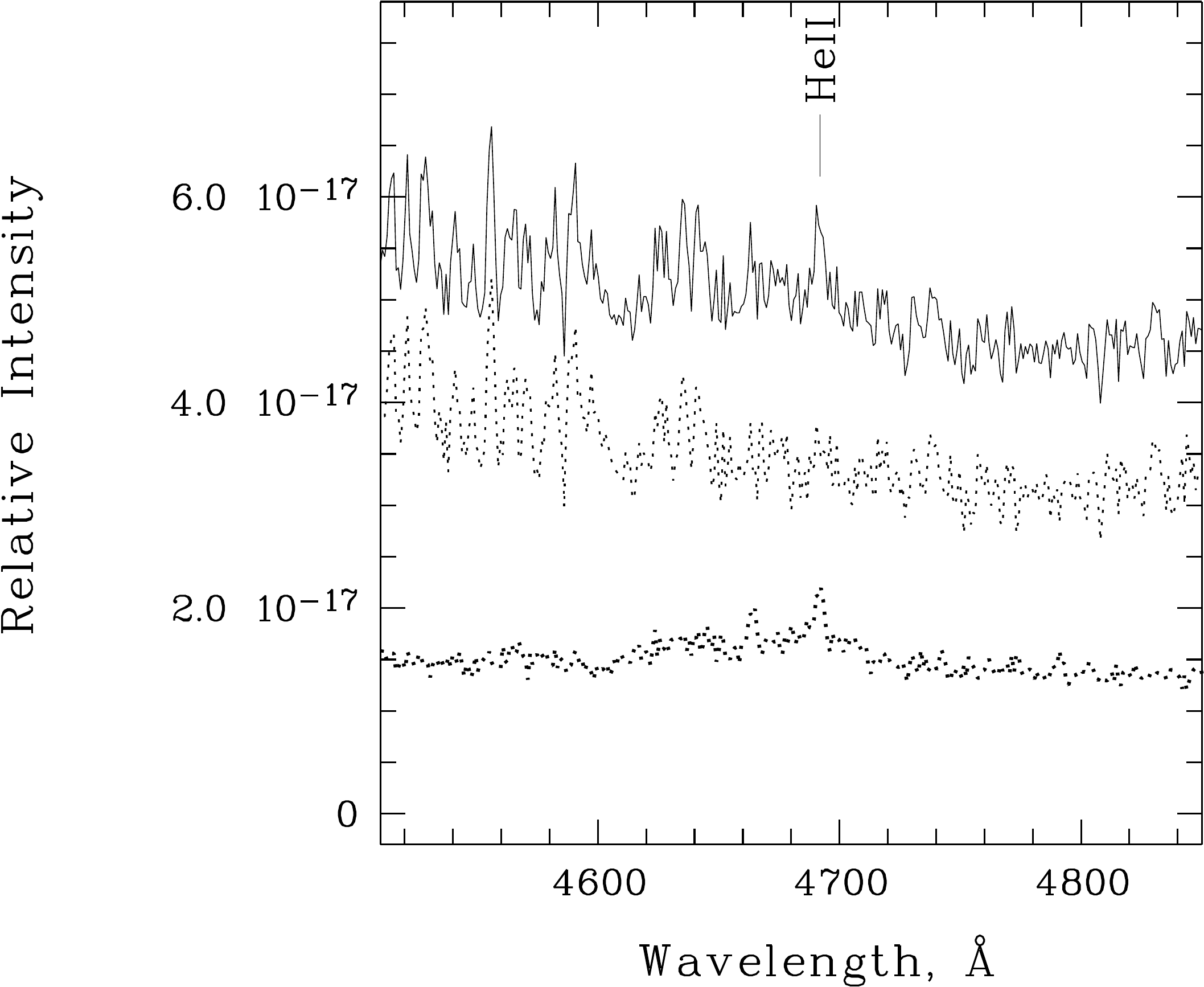}\\
\vspace{1pt}
\caption{
The result of the \ion{He}{ii} line subtraction in the J025941.21+251412.2 spectrum obtained in 2013 (\textit{top panel}) and 2021 (\textit{bottom panel}). The dotted line denotes the spectrum of a neighbouring source (cluster), the solid and dashed lines show the spectrum of J025941.21+251412.2 before and after subtraction of the neighbouring source contribution, respectively. In the top panel, the spectrum of the cluster is shifted upwards by 5.5e-17 for illustrative purposes. In the bottom panel, the subtraction result is shifted down by 2.5e-17.}
\label{Fig5}
\end{center}
\end{figure}

In the 2021 observations, we changed the position angle of the slit to cover (together with the LBV candidate) the nearest star cluster (labelled as <<Star cluster>> in Fig.~\ref{Fig1}) at a distance of about 1\arcsec\ in order to determine the origin of the \ion{He}{ii} line. This cluster, with a brightness comparable to that of J025941.21+251412.2, is one of the most probable places where the \ion{He}{ii} line can be formed. Indeed, analysis of the cluster spectrum revealed \ion{He}{ii} emission, the Bowen blend \ion{C}{iii}+\ion{N}{iii} $\lambda$4650, and \ion{C}{ iv}+\ion{N}{iv} $\lambda$5800 inherent in spectra of Wolf-Rayet stars. Determining the scaling factor based on the intensity of these lines, we subtracted the cluster contribution from the spectrum of the target. This procedure made it possible to completely remove the \ion{He}{ii} line and the Bowen blend from the 2017-2021 spectra of J025941.21+251412.2 (bottom panel of Fig.~\ref{Fig5}). In the case of the 2013 spectrum, depending on the specific multiplying factor, subtraction either leaves a narrow \ion{He}{ii} emission component or removes the line completely together with significantly over-subtracting the broad helium and \ion{C}{iii}+\ion{N}{iii} line components, as is shown in the top panel of Fig.~\ref{Fig5}. This result suggests that in the 2013 data the \ion{He}{ii} line is emitted directly by the LBV candidate, indicating higher temperatures of the stellar atmosphere than in subsequent years.

\begin{figure*}
\begin{center}
\includegraphics[angle=270, width=0.80\linewidth]{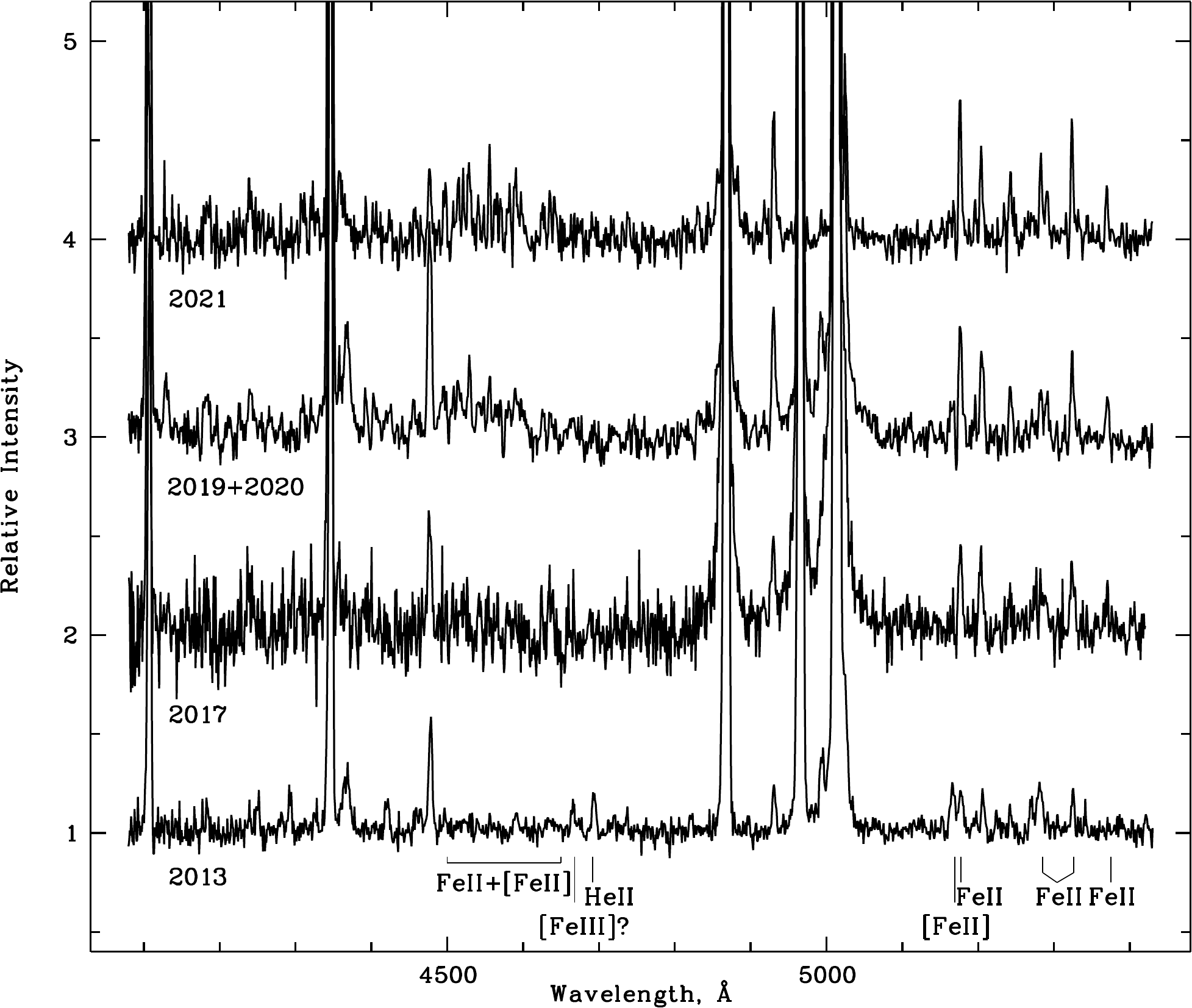}\\
\vspace{5pt}
\includegraphics[angle=270, width=0.8\linewidth]{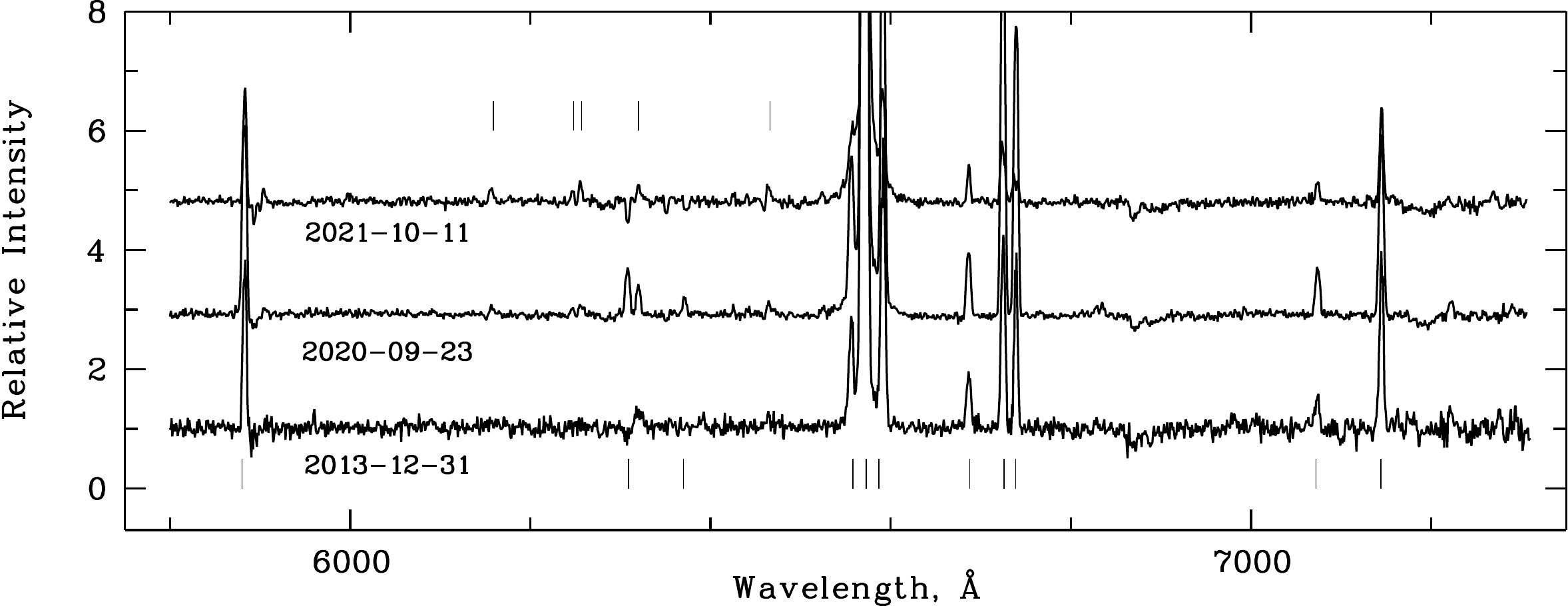}\\
\vspace{1pt}
\caption{
Normalised spectra of J025941.21+251412.2 obtained in different years. 
The top panel shows the spectral lines that undergo the most noticeable changes. In the bottom panel, the unmarked ticks above the spectra represent the \ion{Fe}{ii} lines. The unmarked ticks under the spectra indicate the main lines of the nebula around the object. The nebular line [\ion{O}{iii}] $\lambda$6300 was oversubtracted in the spectra taken in 2013 and 2021.
}
\label{Fig6}
\end{center}
\end{figure*}

The 2013--2021 spectra in the blue and red wavelength ranges obtained as a result of the optimal subtraction of the cluster are presented in Fig.~\ref{Fig6}. The line profiles and intensities show significant variation between the observations. In the 2013 spectra, the hydrogen lines have only narrow components, and there are no iron lines in the $\sim4500-4600$~\AA\AA\ range. The intensities of the other iron lines are much lower, their ratio differs significantly from that in later years. Some \ion{Fe}{ii} lines showing a P Cyg profile with a deep absorption component in the 2021 spectrum are also absent in the 2013 data. The spectra of 2017--2020 are intermediate with respect to the described above. Since the spectra obtained in 2019 and 2020 with the same VPHG1200B grating have no visible difference, in Fig.~\ref{Fig6} we show their sum. Note that the apparent light scattering in the brightest oxygen line [\ion{O}{iii}]\,$\lambda5007$ is due to the proximity of the accumulated signal to the upper limit of the detector dynamic range.

In addition to the spectral changes, the object also demonstrated brightness variations. From 2013 to 2018, J025941.21+251412.2 became brighter by $\Delta \text{R}_c = 0.84 \pm 0.23^m$, whereas from 2018 to 2021 its brightness remained approximately constant (see Table~\ref{tab:phot}).

\subsubsection{J025941.54+251421.8}
\label{J025941.54+251421.8}

\begin{figure*}
\begin{center}
\includegraphics[angle=270, width=0.75\linewidth]{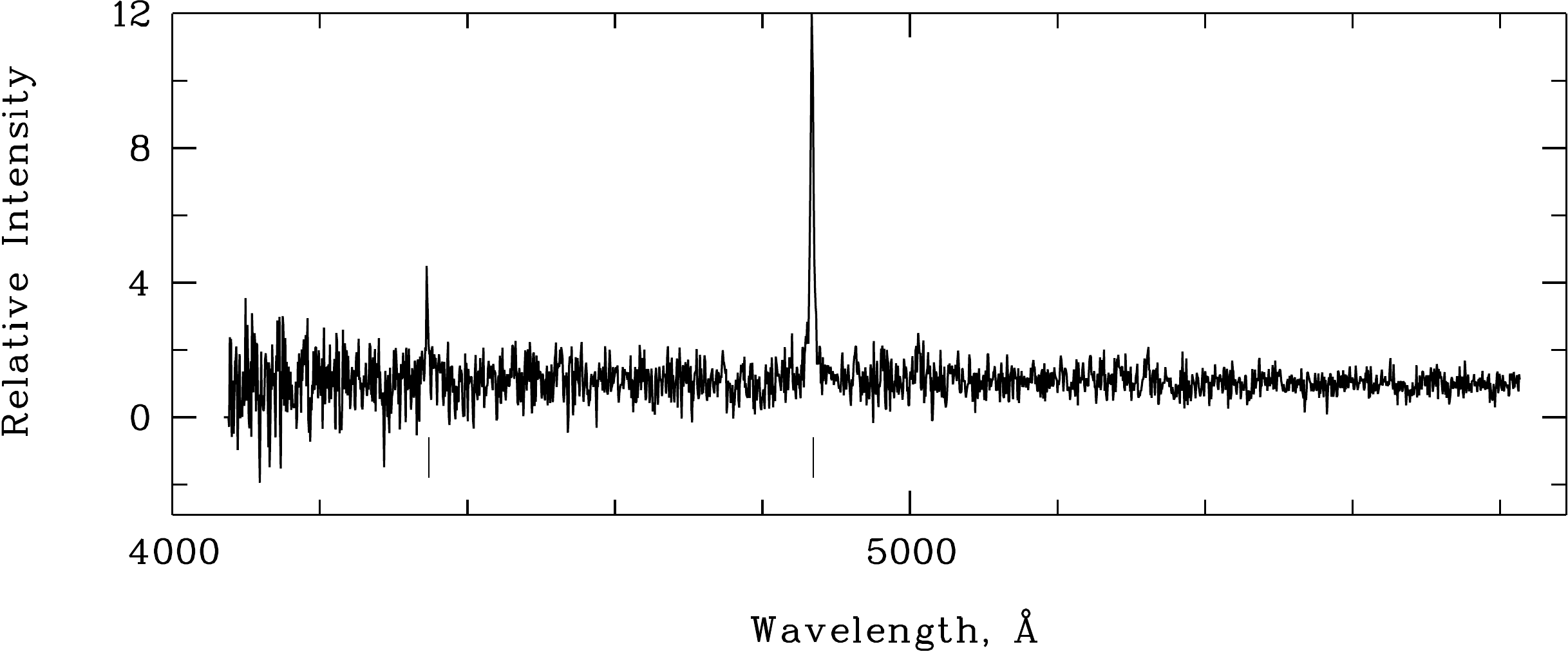}\\
\vspace{5pt}
\includegraphics[angle=270, width=0.75\linewidth]{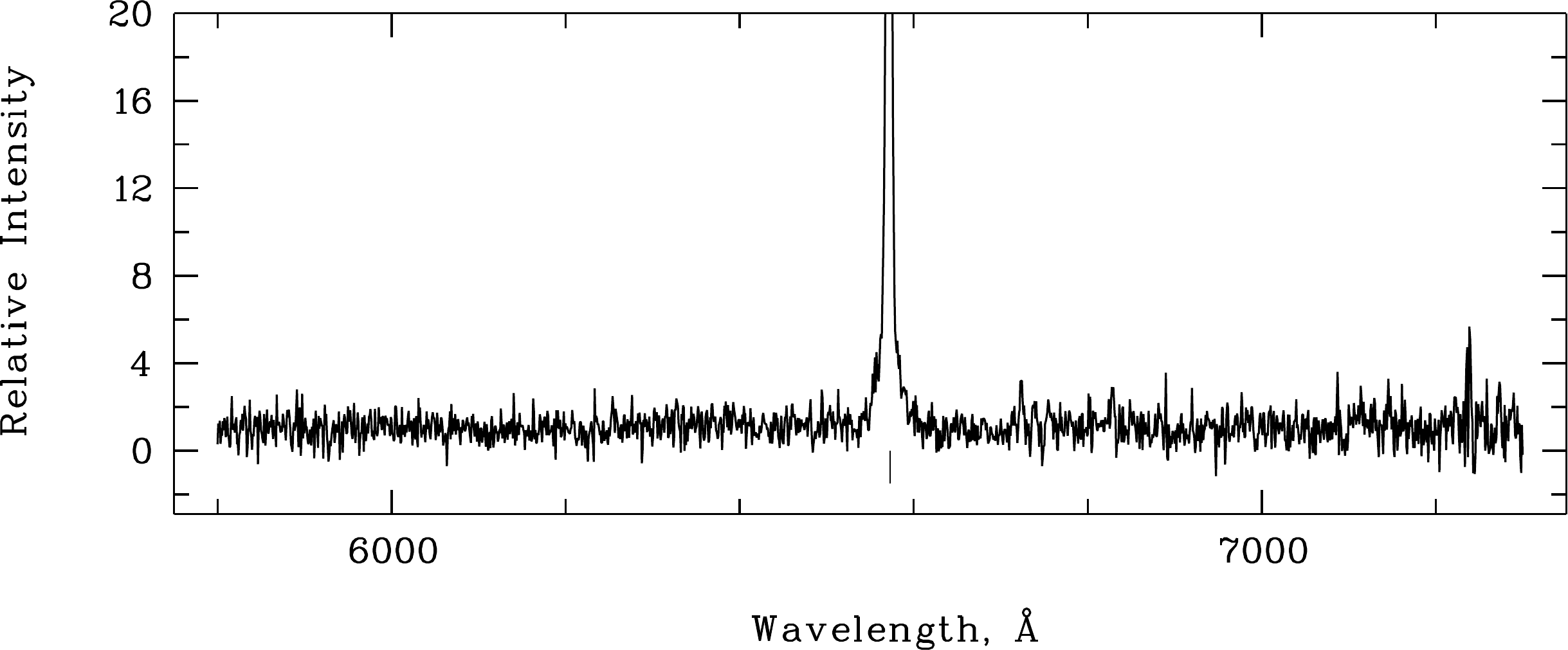}\\
\vspace{1pt}
\caption{Normalised spectrum of J025941.54+251421.8  obtained in 2013. The unlabelled ticks represent the H$\alpha$, H$\beta$ and H$\delta$ lines.}
\label{Fig7}
\end{center}
\end{figure*}

Bright emission features are present only in the spectra of 2013 (Fig.~\ref{Fig7}), when the star was at intermediate brightness with $\text{R} = 21.09 \pm 0.06^m$. One can see in these spectra the hydrogen lines H$\alpha$, H$\beta$ and H$\delta$ with broad wings. The equivalent width of the H$\alpha$ line reaches EW$_{H\alpha}$=$-550\pm120$\AA. The relatively low signal-to-noise ratio of the obtained data does not allow the detection of weaker lines of other elements. In 2020, the magnitude of the object became $\approx22^m$, and the equivalent widths of the hydrogen lines strongly decreased (EW$_{H\alpha}$>-70\AA).

The photometric data of 2005/09/05 hints that the H$\alpha$ equivalent width was the smallest on this date. We found the observed flux in the ACS/HRC/F658N filter to be less than the value derived from the broadband F550M and F814W filters\footnote{The flux conversion between different HST filters was carried out in a way similar to the procedure described above in Sec.~\ref{subsec:imaging} for the conversion between the photometric systems}. This indicates that the line likely had an absorption as well (pure absorption or P Cyg profile). However, the observed and calculated fluxes differ by less than 1-$\sigma$ (of the photometry error of the F658N flux), so we cannot draw an unambiguous conclusion concerning the absence of an emission component. The calculated 3-$\sigma$ equivalent width limit is EW$_{H\alpha}>-4$~\AA.

The total amplitude of the J025941.54+251421.8 brightness decrease is $\Delta \text{R}_c = 2.59 \pm 0.10^m$. In the later observations (2017, 2019 and 2021) taken with the BTA telescope the object became extremely faint, so we were unable to correctly separate it from its relatively bright surroundings, thus these data were not used for photometry.

\subsection{Reddening}

The HST narrow band H$\alpha$+[\ion{N}{ii}] image in the ASC F658N filter (see Fig.~\ref{Fig1}) clearly demonstrates that ionized gas distribution around both LBVs is significantly different. Indeed, there are no any extended emission filaments near J025941.54+251421.8, whereas J025941.21+251412.2 is projected inside the bubble nebulae with the the diameter about 1 arcsec (34 pc). Although we cannot state unequivocally that this system is not a random line of sight projection, it is important to outline the absence of other bright stars near the bubble centre. Therefore this nebulae can be formed by LBV star during its previous evolution and we can use its Balmer decrement to estimate the average $\text{A}_\text{V}$. 

Interstellar reddening was measured using hydrogen line ratios in the spectrum of a nebula taken in several places along the slit between the two studied stars. The ratios of different lines and different places yielded several dozen individual estimates in total. The median value assuming case B photoionisation \citep{Osterbrock06} is $\text{A}_\text{V}=0.90\pm0.13$. This estimate is slightly lower than the average value $\text{A}_\text{V}=1.09$ obtained by \citet{Kim2012}.

\subsection{Stellar parameters}

Since our photometry was performed in a limited number of filters, it was impossible to apply a simple and effective method for estimating the temperature, radius, and bolometric luminosity based on fitting a blackbody model to the star's spectral energy distribution (SED) that we used in our previous works \citep{Solovyeva20, Solovyeva21}. Therefore, the fundamental parameters of J025941.21+251412.2 were determined via modelling its spectrum with the non-LTE code \textsc{cmfgen} \citep{Hillier98}. This code calculates models of stellar atmospheres with a strong gas outflow. We analysed the spectrum with the highest signal-to-noise ratio (BTA/SCORPIO-2, 2021 observations); modelling details will be described in another paper (Solovyeva et al. 2022, in preparation). The best fit \textsc{CMFGEN} model gives a photosphere temperature of T$_{2/3}\approx7900$\,K, bolometric luminosity $\text{L}_{\text{bol}} \approx 1.6 \times 10^6 \text{L}_\odot$ (assuming the measured distance to the galaxy of 7.0~Mpc, see Sec.~\ref{subsec:dist}) and the photosphere radius (at optical depth $\tau = 2/3$) R$_{2/3} \approx 680$~R$_\odot$. We have calculated a grid of models around the optimal one, which made it possible to roughly estimate the parameter errors. The acceptable temperature and luminosity ranges are $7500-8300$\,K and $(1.4-1.8) \times 10^6 \text{L}_\odot$ respectively. The photosphere radius remains almost constant in all the models.
An interesting simulation result is a rather low hydrogen abundance ($\lesssim 20$\%\ by the number of atoms), which indicates the star is highly evolved and on its way to transform into a Wolf-Rayet star.

\begin{figure} 
\center{
\includegraphics[angle=0, width=0.98\linewidth]{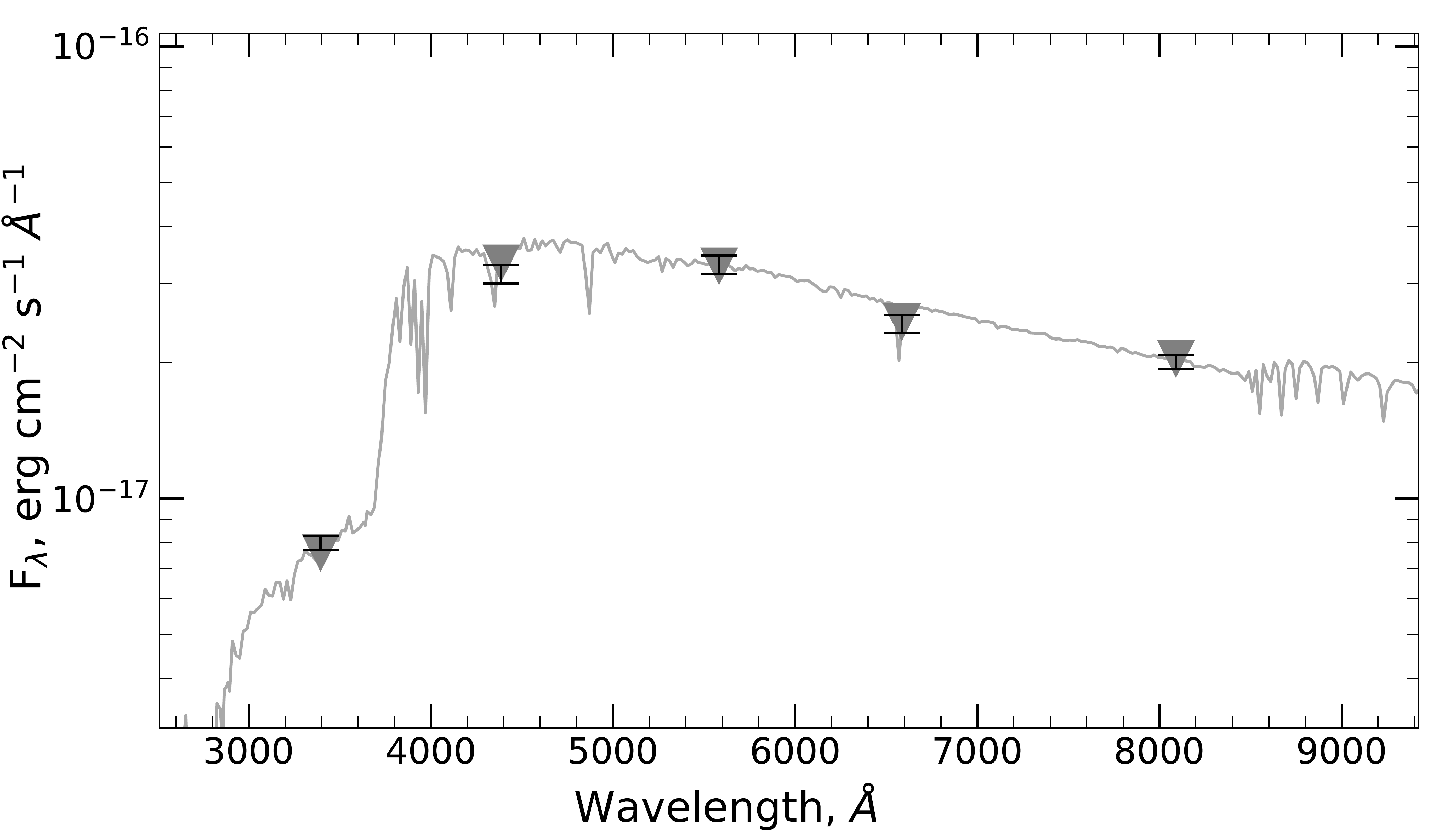}}
\caption{The SED of J025941.54+251421.8. The black solid line represents the best-fit Kurucz model. The black symbols represent the \textit{HST}/ACS/HRC data (2005), the grey triangles show model fluxes in the HST filters.}
\label{Fig8}
\end{figure}

The photosphere temperature of J025941.54+251421.8 was estimated by fitting its SED constructed using the HST/ACS/HRC data (2005). The use of a simple blackbody model gave unsatisfactory results, with interstellar extinction showing significant deviations of the fluxes in the F330W and F658N filters: in both cases, the calculated fluxes turn out to be higher than the observed ones. The flux deficiency at short wavelengths can be interpreted as a rather deep Balmer jump, and in the H$\alpha$ line region it can be associated with the corresponding absorption line (see Sec.~\ref{J025941.54+251421.8} for more details). 
Both features have been perfectly described by Kurucz model atmospheres, which are more advanced than the blackbody approximation. The observed SED was fitted with the models from ATLAS9 \citep{CastelliKurucz2003} using the \textsc{vosa}\,v7.0 \citep{Bayo2008} spectral energy distribution analysis tools. 
The temperature, gravity and interstellar extinction varied in wide intervals: T$_{\text{eff}}=3000-15000$~K, $\log(g)=1-4$ and $\text{A}_\text{V}=0.6-2.5$ (the lower limit corresponds to the Galactic extinction value). The metallicity was fixed at [M/H]=-0.5. The fitting resulted in the parameters T$_{\text{eff}}$=6750\,K and $\log(g)=2.00$ with the 1$\sigma$ errors 330\,K and 0.53, respectively, and a reduced chi-square of $\chi^2$/dof=1.28. The bolometric luminosity, stellar radius and interstellar extinction are $\text{L}_{\text{bol}}=(7.9 \pm 1.5) \times 10^5 \text{L}_\odot$, R$=650\pm90$~R$_\odot$, $\text{A}_\text{V}=0.98 \pm 0.21^m$, which is in good agreement with the estimate from the nebula Balmer decrement. The observed SED with the best-fit model is shown in Fig.~\ref{Fig8}.

\subsection{Age of the environment stars}

\begin{figure} 
\center{
\includegraphics[angle=0, width=0.95\linewidth]{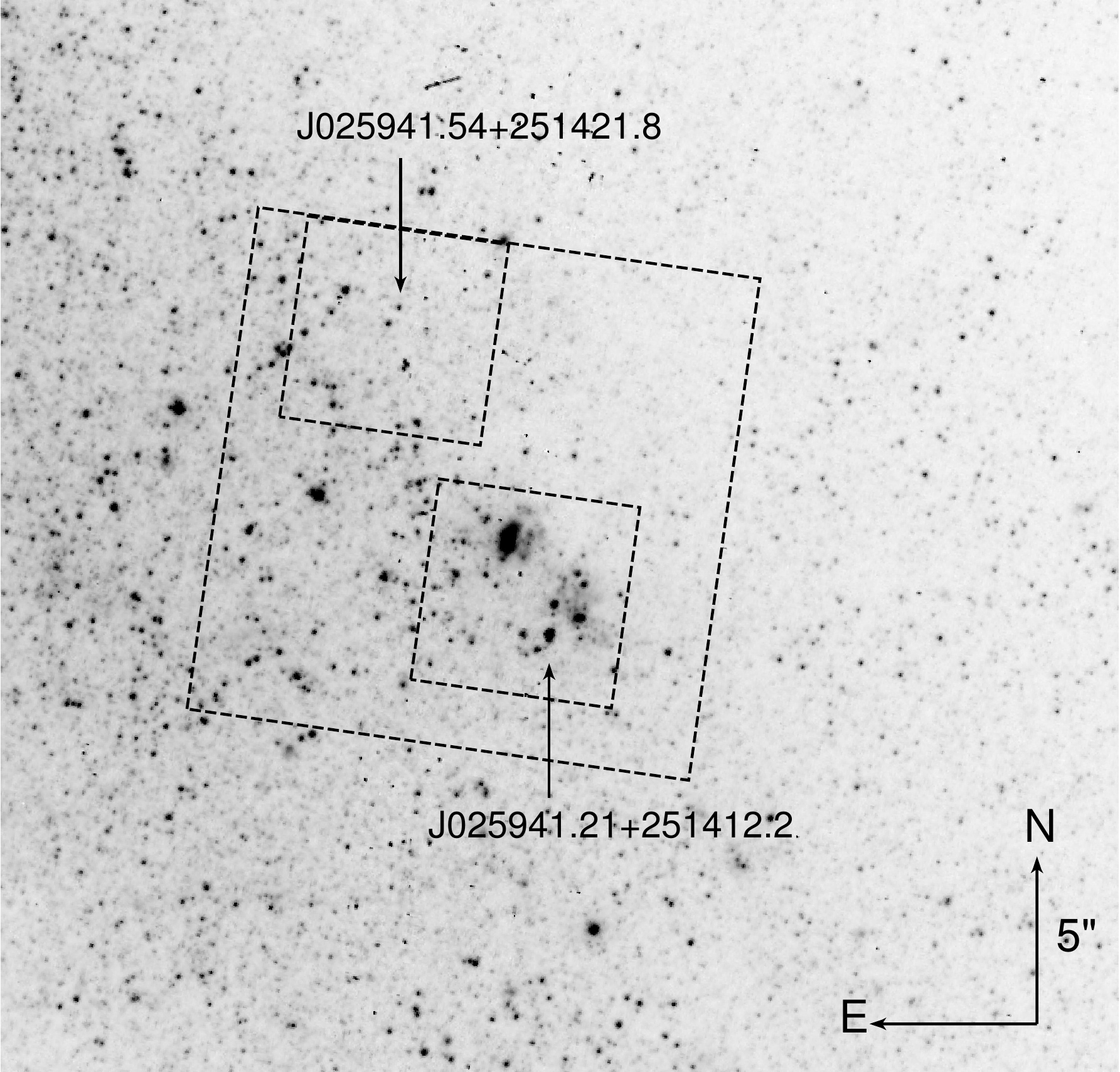}}
\caption{\textit{HST}/ACS/WFC/F814W image with  J025941.21+251412.2 and J025941.54+251421.8 marked by the arrows. The rectangles represent the regions that were used to construct the colour-magnitude diagrams. 
}
\label{Fig9} 
\end{figure}

Both studied objects are located in the region of the galaxy containing a large number of blue stars. Measuring the age of the stellar environment one can restrict the age of our LBV candidates. To estimate the stellar ages, we have constructed a colour--magnitude diagram (CMD) using the results of the PSF photometry from Sec.~\ref{subsec:dist}. We plot on the diagram the stars from the three regions marked in Fig.~\ref{Fig9}: two small 6\arcsec$\times$6\arcsec\ regions in direct vicinity the objects and a larger region (15\arcsec$\times$15\arcsec) covering the area around both cLBVs with many stars. Different regions were used in order to compare and refine the age estimates.

\begin{figure} 
\center{
\includegraphics[angle=0, width=0.9\linewidth]{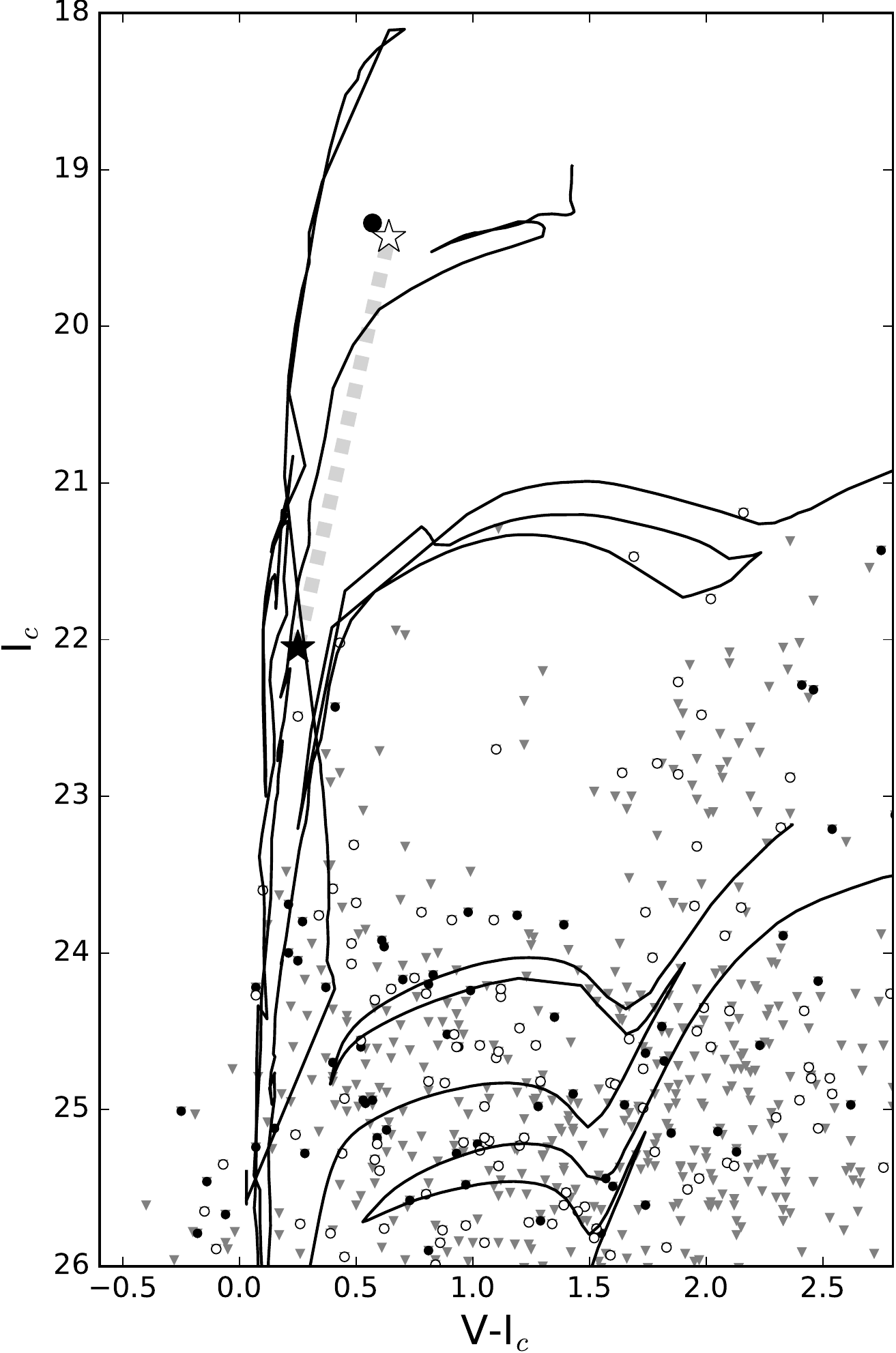}}
\caption{Colour-magnitude diagrams of the stellar environments of the two studied stars. The grey triangles mark the stars of the large star-forming region (Fig.~\ref{Fig9}); the solid and open black circles denote the stars in the small region around J025941.21+251412.2 (big black circle) and J025941.54+251421.8 (star symbol). The grey dashed line denotes the transition of J025941.54+251421.8 between the high (open star) and low (solid star) brightness states. Theoretical isochrones (from top to bottom) are for 3, 5, 10, 50 and 100 Myr.} 
\label{Fig10} 
\end{figure}

Unfortunately, J025941.21+251412.2 fell on a bad pixel column in the first sub-exposure image and on a gap between the CCD chips in the second one in both the HST F606W and F814W data. Therefore, its position on the CMD was determined based on the 2.5-m CMO telescope data of 2018 (V) and 2020 (I$_c$) when the star was at maximum brightness. The magnitude and colour of J025941.54+251421.8 were measured for the high (ACS/HRC, 2005) and low (ACS/WFC, 2019) brightness states. The final CMD diagram with the theoretical isochrones\footnote{The isochrones were obtained from \url{http://stev.oapd.inaf.it/cgi-bin/cmd}} from \citet{Marigo2017} for the metallicity $\text{Z} = 0.008$\footnote{The metallicity value for NGC\,1156 is taken from \citet{Kim2012}} is shown in Fig.~\ref{Fig10}. The isochrones were corrected for interstellar extinction $\text{A}_\text{V} \approx 0.9$ measured above.

The age of the stars on the CMD varies within a fairly wide range. The oldest stars with an age of tens of Myr could have appeared as a result of previous star-formation bursts. A significant part of the red giants are background stars, and they belong to a population not associated with this star forming region. The position of the youngest stars in the three chosen regions corresponds to an isochrone of 10 Myr.

The star cluster with emission features of WR stars near J025941.21+251412.2 (see above) may be used as an additional marker of the age of the stellar environment. The age of such clusters can be fairly accurately estimated using the equivalent width of \ion{He}{ii}~$\lambda$4686, which is about $-6$\AA\ in our case. Based on the diagram of the equivalent width of \ion{He}{ii}~$\lambda$4686 versus the cluster age for $\text{Z} = 0.008$ presented on the Starburst99 web page\footnote{\url{https://www .stsci.edu/science/starburst99/docs/}} \citep{Leitherer1999}, we conclude that the age of this cluster is about 4-5 Myr, which is half of the age determined from the CMD.

\section{Discussion}

As noted above, the spectrum of J025941.21+251412.2 exhibited significant changes from 2013 to 2021. In particular, the bright \ion{He}{ii} $\lambda$4686 emission line observed in 2013 is not visible in the later spectra, while the \ion{Fe}{ii} and [\ion{Fe}{ii}] line profiles became P\,Cyg-like. Such spectral changes accompanied by the photometric variability $\Delta \text{R}_c = 0.84 \pm 0.23^m$ are characteristic only of LBV stars and correspond to the S\,Dor type variability cycle. Thus, taking into account the high bolometric luminosity of $\approx1.6\times10^6$~L$_\odot$ derived from the CMFGEN model at the distance $\approx7.0$~Mpc measured using the TRGB method, we classify J025941.21+251412.2 as an LBV star.

The behaviour of the second discovered star also indicates that it belongs to the LBV type. The variability amplitude of J025941.54+251421.8 was $\Delta \text{R}_c = 2.59 \pm 0.10^m$ from 2004 to 2018. Spectral variability has also been detected in these observations, which is confirmed by the multiple changes in the H$\alpha$ equivalent width from 2005 to 2017 (Sec.~\ref{J025941.54+251421.8}). Moreover, the wide-range ACS/HRC data (2005) show that the H$\alpha$ line most likely was in absorption or had a P Cyg profile, where the absorption component exceeded the emission one. A similar picture can be observed in LBV spectra during their S\,Dor outburts, for example, in the star R71 from the Large Magellanic Cloud during a visual brightening of more than 2$^m$ \citep{Mehner2013, Mehner2017}. During the outburst the photosphere temperature of this LBV was 6650 K, which is below the hydrogen ionisation threshold. At the same time, the radius of the photosphere reached about 600$\text{R}_\odot$ at a bolometric luminosity of $\text{log}(\text{L}/\text{L}_\odot)=5.82$ \citep{Mehner2013}. The parameters of R71 during the outburst are comparable to those of J025941.54+251421.8 in 2005. Thus, we can conclude that J025941.54+251421.8 was experiencing a similar outburst in those observations, but the duration of this outburst is hard to determine due to large gaps in the light curve. The brightness of this star in 2001 was close to that in 2005. In 2013, the brightness became $1.5^m$ fainter and continued to decrease in subsequent years (see Table~\ref{tab:phot}). Thus, assuming that the light curve reflects the change of the J025941.54+251421.8 brightness during one single outburst, we conclude that the eruptive phase of this star lasted for about (or more than) 10 years.

\begin{figure}
\centering
\includegraphics[angle=0, scale=0.45]{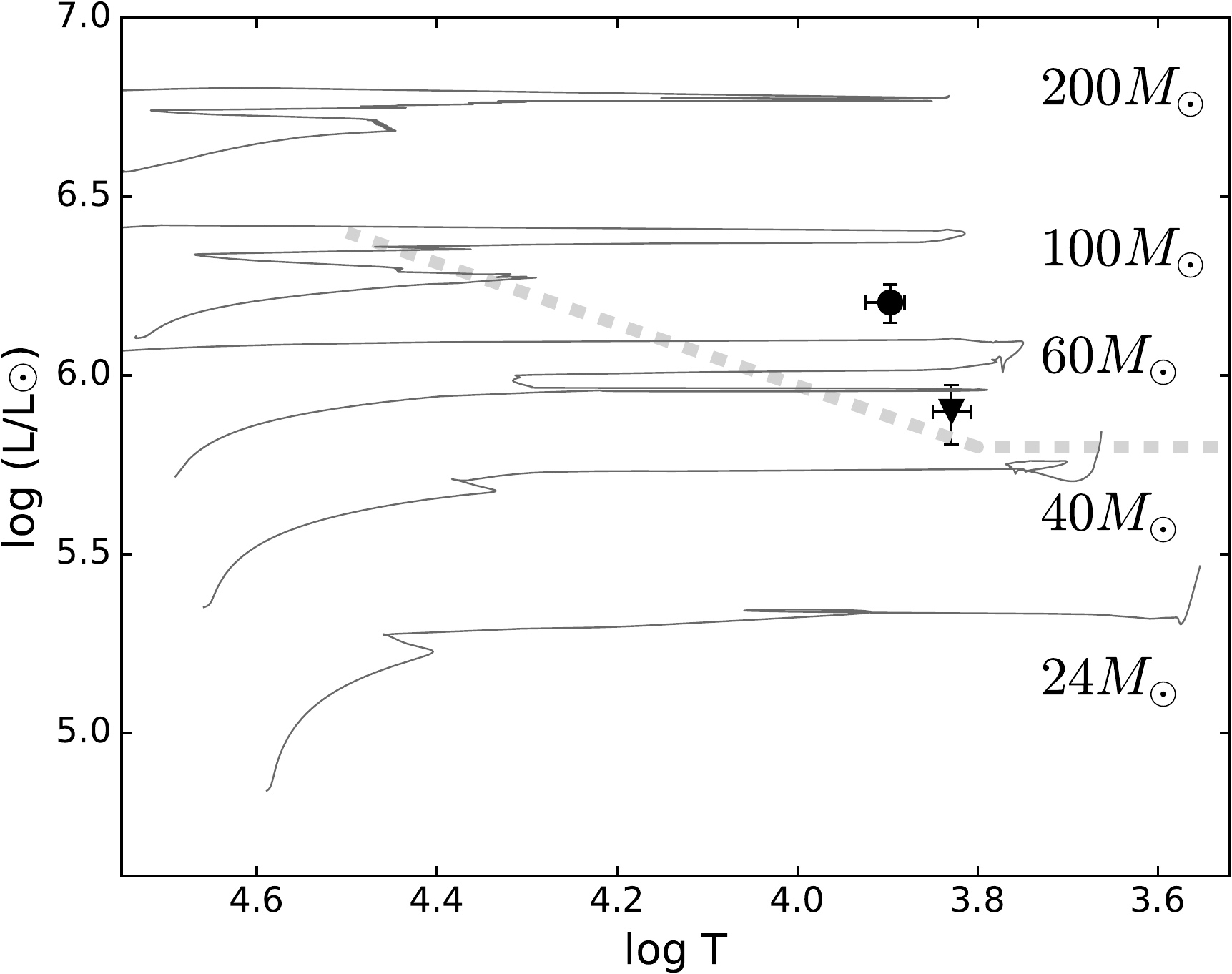}
\caption{Temperature-luminosity diagram with evolutionary tracks of massive stars for a metallicity of $Z=0.008$. Positions of the objects are shown by the circle (J025941.21+251412.2) and triangle (J025941.54+251421.8). Both stars are above the Humphreys-Davidson limit \citep{Humphreys1979} shown by a thick grey line.
}
\label{Fig11}
\end{figure}

We have estimated the initial masses of J025941.21+251412.2 and J025941.54+251421.8 by comparing their positions on the temperature-luminosity diagram with the evolutionary tracks of massive stars assuming a single star evolution scenario (Fig.~ \ref{Fig11}). Using the bolometric luminosities and photosphere temperatures from the modelling above and the evolution tracks for $\text{Z} = 0.008$ from \citep{Tang14}, we have obtained rough mass estimates close to 100~M$_\odot$ for the first star and about 50~M$_\odot$ for the second.

Both mass values can be considered as upper limits because the used bolometric luminosities were obtained in the active state. It is known that $\text{L}_{\text{bol}}$ can sometimes increase by 0.2\,dex compared to the quiescent state even during S\,Dor outbursts. Examples of this are R71 \citep{Mehner2017} and a number of other LBVs \citep{Lamers95, Groh09}. However, we still cannot rule out the giant eruption with shell ejection that is characteristic for SN imposters as the reason for the high variability amplitude of J025941.54+251421.8 (almost 2.6$^m$) and its extremely low photosphere temperature ($\lesssim 7000$\,K) in the bright state. In this case the increase in bolometric luminosity can be more significant compared to the S\,Dor outburst, which has to lead to an even greater overestimation of the stellar mass when comparing the object position on the temperature-luminosity diagram with the evolutionary tracks. Also, given the low brightness of J025941.54+251421.8 in the quiescent state, it cannot be ruled out that it belongs to the class of low-luminosity LBVs with masses around 25~M$_\odot$ \citep{Humphreys16}.

A star of about 100~M$_\odot$ should have a lifetime of slightly more than 3 Myr which is three times less than the age of the stars surrounding J025941.21+251412.2 measured by the isochrone method. However, it is quite close to the age of the nearest cluster with Wolf-Rayet stars (4--5 Myr).
In our previous works we have already reported an age discrepancy between LBVs (and cLBVs) and their surrounding stars \citep{Solovyeva20,Solovyeva21}. This discrepancy is likely associated either with an overestimation of the ages due to a small number of bright stars in the surrounding O-B star associations and clusters, or with the processes of ``rejuvenation'' as a result of mass transfer in close binary systems-- the possible progenitors of LBVs. The discovery of a much younger WR star cluster in this paper supports the first interpretation. On the other hand, the diagram shows traces of several star formation episodes which may imply that we cannot see the youngest stars (with rare exceptions) due to their localisation in small regions around young cluster cores not resolvable into stars even with HST. The second star~--- J025941.54+251421.8~--- with a mass of 25 to $\lesssim50$ solar masses will have an age ranging from $\gtrsim4.6$ to 8 million years towards the final stages of its evolution. Thus, the upper age limit of J025941.54+251421.8 roughly corresponds to the estimated age of the surrounding stars and does not require a ``rejuvenation''.
One of the observational manifestations of LBV formation in the close binary evolution scenario is the relative isolation of some LBVs from O stars \citep{Smith15,Smith16}. For example, according to \citet{Smith15}, there are no O~stars within a radius of more than 250 pc around the star R\,71 \citep{Mehner2017} mentioned above. However, both LBVs in NGC\,1156 are located in a rather dense environment of such stars. Even J025941.54+251421.8, which is more distant from the groups of young stars, is located at a projected distance of less than 60~pc (1.6\arcsec\ at a distance of 7.0 Mpc) from the nearest O-star candidate\footnote{Classified based on a comparison of the tables in \cite{Fitzgerald1970} and \cite{Straizys1981} and the observed $(B-V)_0$ colours, and the absolute magnitude M$_V$ corrected for the interstellar reddening A$_V\simeq1.0$ determined above.}. Thus, we do not yet have reliable evidence for the discovered LBVs being the product of binary evolution.

\section*{Acknowledgements}
Part of the observed data were obtained with a unique scientific facility, the Big Telescope Alt-azimuthal of SAO RAS; data processing and modelling of low resolution spectra were supported under the Ministry of Science and Higher Education of the Russian Federation grant 075-15-2022-262 (13.MNPMU.21.0003). This publication makes use of VOSA, developed under the Spanish Virtual Observatory project.

\section*{Data Availability}
The data underlying this article will be shared on reasonable request to the corresponding author. 

%%%%%%%%%%%%%%%%%%%%%%%%%%%%%%%%%%%%%%%%%%%%%%%%%%



\bibliographystyle{mnras} \bibliography{bibtexbase.bib}

%\begin{thebibliography}{99}
%\end{thebibliography}

%%%%%%%%%%%%%%%%%%%%%%%%%%%%%%%%%%%%%%%%%%%%%%%%%%

%%%%%%%%%%%%%%%%% APPENDICES %%%%%%%%%%%%%%%%%%%%%

% \appendix

% \section{Some extra material}

% If you want to present additional material which would interrupt the flow of the main paper,
% it can be placed in an Appendix which appears after the list of references.

%%%%%%%%%%%%%%%%%%%%%%%%%%%%%%%%%%%%%%%%%%%%%%%%%%

% Don't change these lines
\bsp	\label{lastpage} \end{document}